\documentclass[aps,prl,onecolumn,superscriptaddress,nofootinbib,balancelastpage]{revtex4}

\usepackage{epsfig,amssymb,amsmath,bm}
\usepackage{graphicx}
\usepackage{color}

\def\bea{\begin{eqnarray}}
\def\eea{\end{eqnarray}}
\def\be{\begin{equation}}
\def\ee{\end{equation}}
\newcommand{\de}{\mathrm d}

\newcommand{\om}{{\Omega_{\rm m}}}

\newcommand{\ho}{{H_0}}

\begin{document}
\title{Particle dark matter searches in the anisotropic sky}
\author{Nicolao Fornengo}
\author{Marco Regis}
\affiliation{\rm Dipartimento di Fisica, Universit\`a  di Torino and INFN, Torino, Italy \\
fornengo@to.infn.it, regis@to.infn.it}

%\date{\today}

\begin{abstract}
Anisotropies in the electromagnetic emission
produced by dark matter annihilation or decay in the extragalactic sky are a recent tool in the quest for a particle dark matter evidence.
We review the formalism to compute the two-point angular power spectrum in the halo-model approach and
discuss the features and the relative size of the various auto- and cross-correlation signals 
that can be envisaged for anisotropy studies. 
From the side of particle dark matter signals, we consider the full multi-wavelength
spectrum, from the radio emission to X-ray and gamma-ray productions. 
We discuss the angular power spectra of the auto-correlation of each of these signals and of the cross-correlation
between any pair of them. We then extend the search to comprise specific gravitational
tracers of dark matter distribution in the Universe: weak-lensing cosmic shear, large-scale-structure matter distribution and CMB-lensing. 
We have shown that cross-correlating a multi-wavelength dark matter signal
(which is a direct manifestation of its particle physics nature) with a gravitational tracer (which is a manifestation of the presence of large amounts of unseen matter in the Universe) 
may offer a promising tool to demonstrate that what we call DM is indeed formed by elementary particles.
\end{abstract}

\maketitle

\section{Introduction}

The presence of a large amount of dark matter (DM) in the Universe is now supported by
means of a numerous and converging astrophysical and cosmological probes. These probes allow to
determine that the average, horizon-scale, DM contribution to the energy density of the Universe is
six times larger than the baryonic density.
In a spatially flat $\Lambda$CDM scenario the DM energy density is about
25\% of the critical density of the Universe \cite{Ade:2013zuv}. 
On smaller scales, the DM is observed to be distributed quite anisotropically: 
it forms a hierarchical network of cosmic structures, from
the large-scale galaxy clusters down to the small-scale galaxies and their inner parts.
This observational picture is consistent with the theoretical understanding of 
cosmic structure formation through gravitational instability, based on a DM component which
is relatively cold. The cold dark matter (CDM) paradigm, with CDM evolving in an expanding Universe 
%dominated at ``recent" times by a cosmological-constant-like dark energy (the $\Lambda$CDM model) 
is able to reproduce
the statistical properties of the large scale structure (LSS) of the Universe, and predicts a galaxy
formation pattern which is consistent, at least to a large extent, to observations.
Although the common picture of CDM structure formation may be posed under discussion at
very small scales, and even though the true role of baryons if the galaxy formation process has
only recently started to be investigated, nevertheless the general picture of the LSS formation
is robust (for recent reviews on the subject, see, e.g., \cite{Bertone:2010zza}).

The most common interpretation of the DM in the Universe relies on the simple and quite natural
paradigm of  particle excitation in the hot early plasma, with subsequent decoupling and formation
of a cosmological abundance of DM particles. CDM is then naturally explained in terms of relic particles
which need to be relatively massive (in order to be cold at decoupling) and weakly interacting (such that they freeze-out
with the correct relic abundance). Although alternative hypothesis are possible (axions or axion-like
particles, non-thermal or non-symmetric DM production) the WIMP mechanism offers a viable and
natural solution to the presence of DM in the Universe.

DM evidence is currently of pure gravitational origin: galaxy-cluster dynamics, rotational curves
of spiral galaxies, gravitational lensing observations, hydrodynamical equilibrium of hot gas in galaxy clusters, the same energy budget of
the Universe and the theory of structure formation, are all probes based on the gravitational effects
induced by the presence of large amounts of DM. However, if DM is composed by elementary particles,
vestiges of the early thermal phase, it is expected to produce also non-gravitational signals, specifically
related to its particle-physics nature. A large number of astrophysical DM signals are currently
explored: except for direct detection, which relies on the direct scattering of the DM particles with a
low-background detector in an underground laboratory, all other foreseen signals are of indirect
origin and rely on the products of annihilation (or decay) of the DM particle in the astrophysical
environment, either galactic or extra-galactic. These indirect signals comprise charged galactic exotic cosmic-rays species (electrons/positrons, antiprotons, antideuterons), the whole spectrum of electromagnetic emission (from radio
to gamma-rays) and neutrinos. 

The indirect detection signals are intrinsically anisotropic, since they are produced 
by DM structures in the Universe, and structures are present potentially at any
scale: galaxy clusters on the largest scales, individual galaxies inside clusters, 
subhalos inside galaxies. Especially electromagnetic signals, which more directly trace the distribution of the DM that produces them, are expected to exhibit some level of anisotropy. Bright enough DM objects (at any given wavelength) would appear as resolved sources: an example could be a radio or gamma-ray
halo surrounding and comprising a galaxy cluster or a single galaxy (dwarf galaxies, being DM dominated objects,
are interesting targets for this search) or even galactic subhalos which could appear as isolated objects
in the radio or gamma-rays sky).
Unresolved DM sources, i.e. with brightnesses below the detector sensitivity,  produce instead
a diffuse flux of photons which contains a subdominant level of anisotropy, due to the sum of a large number of
faint sources related to the anisotropic DM distribution. In this case, the effect of anisotropies could be identified on a statistical basis, by means of studying observables related to angular correlations
of the spatial map of the signal, the most typical and relevant one being the two-point angular power spectrum.
We comment that also the diffuse neutrino signal would be endowed of the anisotropic feature, since
neutrinos are necessarily produced in the same processes that lead also to prompt gamma-rays:
however, the mechanisms of neutrino detection introduce a larger spatial uncertainty, as compared
to photon detectors, and moreover the detection rates for neutrinos are significantly lower than
for photons, due to the weak nature of neutrino interactions.

Indirect detection searches of the anisotropic DM sky can follow three
directions: the study of anisotropies of a single electromagnetic signal, which
represents an auto-correlation observable; the cross-correlation of anisotropies in two different
electromagnetic signals; the cross-correlation of an electromagnetic signal with gravitational
tracers of the DM distribution in the Universe. 
The first option has already been attempted both for gamma-rays \cite{Ando:2005xg,Ando:2006cr,Fornasa:2009qh,Ibarra:2009nw,Cuoco:2010jb,Fornasa:2012gu,Hensley:2012xj,Ripken:2012db,Ando:2013ff} (also on the data recently reported by the Fermi Collaboration~\cite{Ackermann:2012uf}) and for the radio signal \cite{Zhang:2008rs,Fornengo:2011xk} (for some discussions in the X-ray band, see \cite{Inoue:2013vza}).

Different wavelengths are emitted by DM through different physical mechanisms: gamma-rays
may be promptly produced in the DM annihilation/decay event through production of mesons
(mostly neutral pions) or baryons, which subsequently decay, or as final-state radiation of charged particles; X-rays (and gamma-rays) 
are produced by inverse-Compton scattering of the electrons/positrons
produced by DM annihilation/decay on the interstellar radiation fields and on the CMB; radio waves are
due to synchrotron emission of the same electrons/positrons in the magnetic fields present in the
cosmic structure where they are produced. The different electromagnetic signal therefore probe
differently the astrophysical properties of the emitting DM structure. 
Prompt gamma-ray emissions most directly retain the spatial features of the DM density field, while the radiative signals feel also the
electrons/positrons spatial diffusion and energy loss, as well as the distribution of magnetic fields. On the other hand, the angular resolution of gamma-rays detectors is lesser than the one accessible to radio (and X-ray) telescopes. Each signal has therefore different features, and the possibility to exploit all types of
signal can potentially help in extracting a true DM signal. 

Let us notice here that the relative size
of the electromagnetic emission at different wavelengths from DM particles, can significantly change
depending on the particle DM properties. For example, a DM particle that can annihilate into hadrons,
typically produces softer spectra both of gamma-rays and of electrons/positrons, as compared to
leptonic annihilations: this reflects on the spectral features of both the prompt and radiative signals,
and to some extent also on the spatial features of the radiative signal, since diffusion and energy losses
of electrons/positrons depend on energy. Similar arguments apply to the case of decaying DM.
The spectral features then depend also on the DM mass, since non-relativistic annihilations (or decays)
set an upper bound on the maximal accessible energy at the value of the DM mass (or half the value
of the DM mass for DM decay). All these features can then affect differently the relative size 
also of the auto-correlation signals.

The second possibility, i.e. cross-correlation between different electromagnetic signal, has not yet been studied: by employing the signal coming at two different wavelengths, we can probe in an independent and richer
way the different responses of a DM signal to the DM particle properties (as discussed above), and to some extent also to the properties of the DM structures (again, for the reasons discussed above on 
the way an electromagnetic signal is produced). In the following we will therefore discuss the potentialities for a DM signal in the different cross-correlation channels and try to quantify the mutual impact in the signal extraction. We will explicitly discuss radio-gamma, radio-X and X-gamma
cross-correlations.

The third option is potentially the most relevant one, since it attempts to directly correlate
two distinctive features of particle DM: an electromagnetic signal, which is a typical and
unambiguous
manifestation of the DM being an elementary particle, from one side; the direct probe of the existence of
DM in the Universe, which is seen through the gravitational effects induced by the presence
of DM in the Universe, on the other side. A positive signal in this cross-correlation channel would provide
direct evidence that what is measured by means of gravitational probes is indeed due to DM in terms of an elementary particle and is not, e.g., a manifestation of alternative theories of gravity. 
This third approach has been pioneered in Ref.~\cite{Camera:2012cj}, where the cross-correlation of the 
gamma-rays emission with the weak-lensing cosmic shear has been studied and shown to
represent a viable and promising novel channel of investigation. In this paper we extend the
cross-correlation studies to two additional gravitational probes: tracers of the large scale structure and the very recent observation of the CMB lensing signal. We also extend the
analyses to comprise not only gamma-rays, but also the radio and X-rays emissions from DM,
in order to establish the potentialities of the full range of possibilities.

In this paper we concentrate on the theory of extragalactic DM anisotropies.
We discuss the features, sizes and comparisons of the various auto- and cross-correlation signals 
in the anisotropic sky, for DM annihilations and decays and looking into the whole electromagnetic spectrum, 
with the aim of exploring the full range of possible signals and correlations, and of setting the ground for future analyses. 
First, we will carefully outline the formalism needed to perform computations of extragalactic DM anisotropies.
We then review some channels already discussed in the literature (gamma-rays autocorrelations, radio autocorrelations and gamma rays/cosmic shear cross-correlation), and extend the range of investigation to propose new
channels: for the signals related to the particle physics nature of DM, we explicitly discuss the auto- and mutual cross-correlations in the whole
electromagnetic spectrum, by adding to the study the gamma/radio case and the auto- and cross-correlation signals involving X-rays; for the signals able to correlate the gravitational probes
of DM in the Universe to the particle-physics ability to produce photon fluxes,
we extend the idea to investigate the gamma-rays/cosmic-shear correlation, originally proposed in Ref. \cite{Camera:2012cj},
to further adopt radio and X-rays emissions from one side, and to embrace
large-scale-structure tracers \cite{Ando:2013xwa} and CMB lensing from
the other side.
A second step of this complex type of analysis will be a detailed study of the different, various
and intricate astrophysical emitters at the different wavelengths (like active galactic nuclei, star forming galaxies, etc.) 
which also contribute to the anisotropic sky and represent the background over which the
signal we are discussing have to be confronted, as well as an accurate assessment of the observational capabilities. 
%We will comment about these sources of background, but 
We leave these parts of the analysis to future works.

The paper starts  with a review of the general formalism apt to derive the two-point angular power spectrum
of correlation signals, and details the steps relevant for studying both DM annihilation (which depends
on the square of the DM density field) and DM decay (which instead depends linearly on the DM density) cases. 
Then discusses the features of the relevant ingredients which enter the correlation
signal: the three-dimensional power spectrum, both for the auto- and cross-correlation, and the 
window functions for each electromagnetic and gravitational signal. 
Angular power spectrum predictions are then shown and discussed for all types of correlations signals: auto correlations in the radio, X and
gamma-rays bands; cross-correlations among them; 
cross-correlations between the radio or gamma-rays emission and the gravitational probes (cosmic shear, CMB lensing, LSS observables). 
Results will be presented for benchmark cases and for both annihilating and decaying DM.

\section{General Formalism}

In this Section we review the general formalism to derive the angular power spectrum of the auto- and cross-correlation of any given pair of intensity fields, which in our analysis are
given by the radio, X-ray, gamma-ray DM fluxes, or the cosmic shear, LSS, CMB-lensing maps.

The source intensity $I_g$ along a given direction $\vec n$ can be written as:
\be
 I_g (\vec n) = \int d\chi\, g(\chi,\vec n)\,\tilde W(\chi)
  \label{eq:int}
\ee
where $\chi(z)$ denotes the radial comoving distance (function of the redshift $z$), $g(\chi,\vec n)$ is the density field of 
the source, and $\tilde W(\chi)$ is the window function (which is function of the distance but does not depend on the specific direction
$\vec n$).
We can also define a normalized version of the window function as $W(\chi)=\langle g \rangle\,\tilde W(\chi)$, such that the average intensity is obtained by integration of
the normalized window function along the distance:
$\langle I_g \rangle = \int d\chi\ W(\chi)$.

The window function brings the information on how the actual
observable related of the signal under study is distributed in redshift.
It will act as a weight for the different redshift contributions of the 3D 
power spectrum when computing the angular power spectrum.
Its specific form varies depending on whether we are studying the cosmological emission of an
electromagnetic signal from DM annihilation (or decay), or instead the images of background galaxies (which are lensed by the structures) in the case of cosmic shear, or the CMB flux at the large scattering surface in the case of CMB lensing, or the distribution of astrophysical sources and of their light emission (typically in the visible, infrared or radio wavelength) in the case of LSS tracers. The specific form of $W(\chi)$ for all the cases under study will be reported in a dedicated Section below.

By defining the intensity fluctuation as $\delta I_g(\vec n)\equiv I_g (\vec n) - \langle I_g \rangle$ and expanding the fluctuation field in terms of spherical 
harmonics, $\delta I_g(\vec n) = \langle I_g \rangle \sum_{\ell m} a_{\ell m} Y_{\ell m}(\vec n)$, the (dimensionless) $a_{\ell m}$ coefficients can be expressed as:
\be
 a_{\ell m} =\frac{1}{\langle I_g \rangle} \int\, d\vec n\, \delta I_g(\vec n)\,Y_{\ell m}^\ast (\vec n)
=\frac{1}{\langle I_g \rangle} \int\, d\vec n\, d\chi\ f_g(\chi,\bm r) W(\chi)\,Y_{\ell m}^\ast (\vec n)
  \label{eq:alm}
\ee
where the orthonormal relations for $Y_{\ell m}$ have been used and where $f_g \equiv g/\langle g \rangle-1$. A spatial Fourier transformation of $f_g$ and the use of the Rayleigh expansion
of a plane wave into spherical harmonics and spherical Bessel functions $j_{\ell}$, allows to
derive:
\bea
 a_{\ell m} &=& \frac{1}{\langle I_g \rangle}\int\,d\vec n\,d\chi\,\frac{d\bm k}{(2\pi)^3} \hat f_g(\chi,\bm k) e^{i\bm k\cdot \bm r}W(\chi)Y_{\ell m}^\ast (\vec n)\nonumber\\
 &=& \frac{1}{\langle I_g \rangle}\int\,d\vec n\,d\chi\,\frac{d\bm k}{2\pi^2} \hat f_g(\chi,\bm k) \left[ \sum_{\ell^\prime m^\prime} i^{\ell^\prime} j_{\ell^\prime} (k\chi) Y_{\ell^\prime m^\prime}^\ast (\hat{\bm k}) Y_{\ell^\prime m^\prime} (\vec n) \right]W(\chi)Y_{\ell m}^\ast (\vec n)\nonumber\\
 &=& \frac{i^\ell}{\langle I_g \rangle} \int d\chi\ W(\chi) \int \frac{d\bm k}{2\pi^2}\
  \hat f_g(\chi,\bm k) j_\ell(k\chi) Y_{\ell m}^\ast (\hat{\bm k})
  \label{eq:alm2}
\eea
where we used $\vec r =\chi\,\vec n$.

The angular power spectrum (PS) is defined as 
$C_\ell^{(ij)} = \langle \sum_m a_{\ell m}^{(i)} a_{\ell m}^{(j)*}\rangle$, where 
$i,j=1,2$ label the two signals for which the correlation is studied (e.g., $i=$ gamma-rays and $j=$ cosmic-shear), and the brackets $\langle \dots \rangle$ denote ensemble average.
Clearly, the cross-correlation angular PS $C_\ell^{(ij)}=C_\ell^{(ji)}$, as can be easily derived from the relation
$a_{\ell }=(-1)^m a_{\ell -m}^*$, while auto-correlation occurs for $i=j$. By using
Eq.~(\ref{eq:alm2}), the correlation angular PS takes the form:
\bea
 C_\ell^{(ij)} &=& \frac{1}{\langle I_i \rangle\langle I_j \rangle} \int d\chi\ W_i(\chi) \int d\chi^\prime\ W_j(\chi^\prime)
  \int \frac{d\bm k}{2\pi^2} \int \frac{d\bm k^\prime}{2\pi^2}\  \langle \hat f_{g_i}(\chi,\bm k)\hat f_{g_j}^\ast (\chi^\prime,\bm k^\prime) \rangle
  j_\ell(kr) j_{\ell^\prime}(k^\prime r^\prime) Y_{\ell m}(\hat{\bm k}) Y_{\ell^\prime m^\prime}^\ast (\hat{\bm k^\prime})
  \nonumber\\
 &=&  \frac{2}{\pi\langle I_i \rangle\langle I_j \rangle} \int d\chi\ W_i(\chi) \int d\chi^\prime\ W_j(\chi^\prime)
  \int d\bm k\ P_{ij}(k,\chi,\chi^\prime)
  j_\ell(kr) j_{\ell^\prime}(k r^\prime) Y_{\ell m}(\hat{\bm k}) Y_{\ell^\prime m^\prime}^\ast (\hat{\bm k})
  \nonumber\\
&=&  \frac{2}{\pi\langle I_i \rangle\langle I_j \rangle} \int d\chi\ W_i(\chi) \int d\chi^\prime\ dk\, k^2\ W_j(\chi^\prime)
 P_{ij}(k,\chi,\chi^\prime) j_\ell(kr) j_\ell(k r^\prime)
  \nonumber\\
&=&  \frac{1}{\langle I_i \rangle\langle I_j \rangle} \int \frac{d\chi}{\chi^2} W_i(\chi)\, W_j(\chi)
 P_{ij}(k=\ell/\chi,\chi)
\label{eq:clfin}
\eea
where in the second step we introduced the definition of the three-dimensional (3D)
power-spectrum $P_{ij}$ through
$\langle \hat f_{g_i} (\chi,\bm k) \hat f_{g_j}^\ast (\chi^\prime,\bm k^\prime) \rangle = (2\pi)^3 \delta^3 (\bm k - \bm k^\prime) P_{ij}(k,\chi,\chi^\prime)$, and in 
the last step we assumed the Limber approximation \cite{limber} to hold for such PS.

The next step is to compute the explicit form for the specific 3D power spectra $P_{ij}$,
which are the Fourier transform of the two-point correlation functions (2PCF) in real space: 
$\xi^{(2)}_{ij}(\bm x, \bm y) \equiv \langle f_{g_i}(\bm x) f_{g_j}(\bm y) \rangle$.
Following Ref.~\cite{Scherrer:1991}, we assume that the density field 
can be expressed as the sum of independent seeds (i.e., of discrete masses,
in the case of the gravitational tracers, or the ensuing electromagnetic emission):
\be
f(\bm x)=\sum_a\,f(m_a,\bm x-\bm x_a)=\int dm \int d^3\bm x' \sum_a\,\delta^3(\bm x'-\bm x_a)\delta(m-m_a) f(m,\bm x-\bm x')
\label{eq:denseed}
\ee
where $a$ labels the seeds, the $\delta$'s are Dirac-delta functions, and we 
take the mass $m$ to be the parameter which characterizes the seeds.
The seed density can be expressed as:
 $dn/dm=\langle \sum_a\, \delta^3(\bm x-\bm x_a)\delta(m-m_a) \rangle$, where
 $\langle \dots \rangle$ denotes the ensemble average over of all possible 
seed distributions. With these definitions, the 2PCF reads: 
{\small
\be
\xi^{(2)}(\bm x ,\bm y)=\int dm_1\,dm_2\, d^3\bm x_1\,d^3\bm x_2 \langle \sum_a\, \delta^3(\bm x_1-\bm x_a)\delta(m_1-m_a)\,\sum_b\, \delta^3(\bm x_2-\bm x_b)\delta(m_2-m_b) \rangle f_1(m_1,\bm x-\bm x_1)\,f_2(m_2,\bm y-\bm x_2)
\ee}
The correlation of a seed $a$ with mass $m_1$ at position $x_1$ with a different 
seed $b$ with mass $m_2$ at position $x_2$ is provided by the seed-2PCF  
$\xi_s^{(2)}(m_1,m_2,\bm x_1 ,\bm x_2)$, and it is not difficult to see that:
{\small
\be
\langle \sum_a\, \delta^3(\bm x_1-\bm x_a)\delta(m_1-m_a)\,\sum_b\, \delta^3(\bm x_2-\bm x_b)\delta(m_2-m_b) \rangle=
\frac{dn}{dm_1}\frac{dn}{dm_2}[1+\xi_s^{(2)}(m_1,m_2,\bm x_1 ,\bm x_2)]+\frac{dn}{dm_1}\delta^3(\bm x_1-\bm x_2)\delta(m_1-m_2)
\label{eq:2corr}
\ee
}
which leads to: 
\bea
\xi^{(2)}(\bm x ,\bm y) &=&\int dm\, d^3\bm x_1 \frac{dn}{dm}f_1(\bm x-\bm x_1,m) f_2(\bm y-\bm x_1,m) \nonumber \\
&+&\int dm_1\,dm_2\, d^3\bm x_1\,d^3\bm x_2 \frac{dn}{dm_1}\frac{dn}{dm_2}f_1(\bm x-\bm x_1,m_1) f_2 (\bm y-\bm x_2,m_2) \xi_s^{(2)}(m_1,m_2,\bm x_1,\bm x_2)
  \label{csi2x}
\eea
Coming back to the fact that the 3D power spectrum $P_{ij}$ is the Fourier transform of $\xi^{(2)}(\bm x ,\bm y)$, and writing $f_i$ in terms of their Fourier transforms $\hat f_i$, one obtains:
\be
 P_{ij}(k) = \int dm\ \frac{dn}{dm} \hat f_i^\ast(k|m)\,\hat f_j(k|m) + \int dm_1\,dm_2\, \frac{dn}{dm_1}\frac{dn}{dm_2} \hat f_i^\ast(k|m_1) \hat f_j(k|m_2)\,P^{(s)}(k,m_1,m_2)\label{eq:PSs}
\ee
where the power spectrum of the seed distribution $P^{(s)}$ is the Fourier transform of 
$\xi_s^{(2)}(m_1,m_2,\bm x_1 ,\bm x_2)$.

In our analysis, we often refer to mass density fluctuations. In this case,
the seed-2PCF is the (homogeneous and isotropic) linear correlation function of matter (notice that it has to be the linear one, since in Eq.~(\ref{eq:denseed}) 
we wrote the density field as a linear superposition of seeds), $\xi_s^{(2)}(m_1,m_2,\bm x_1 ,\bm x_2)=\xi_{\rm lin}^{(2)}(|\bm x_i-\bm x_j|)$.
The corresponding power spectrum is the customary $P^{\rm lin}(k)$.
For other objects considered in the following, i.e. DM halos and astrophysical sources, we will assume (as usually done) that their 2PCF $\xi_s^{(2)}$ can be 
related to the linear correlation function of matter
 by means of $\xi_{s,ij}^{(2)}(m_1,m_2,\bm x_i,\bm x_j) \approx b_i(m_1)\,b_j(m_2)\,\xi_{\rm lin}^{(2)}(|\bm x_i-\bm x_j|)$ where $b_i(m)$ is the linear bias between the object $i$ and matter.
Thus using $P^{(s)}(k,m_1,m_2)=b_i(m_1)\,b_j(m_2)\,P^{\rm lin}(k)$, we finally 
arrive at the decomposition of the 3D power-spectrum into the one-halo and two-halo
terms:
\be
P_{ij}(k) = P_{ij}^{1h}(k) + P_{ij}^{2h}(k)
\ee
where:
\bea
 P_{ij}^{1h}(k) &=& \int dm\ \frac{dn}{dm} \hat f_i^\ast(k|m)\,\hat f_j(k|m) \label{eq:PS1halo}
 \\
 P_{ij}^{2h}(k) &=& \left[\int dm_1\,\frac{dn}{dm_1} b_i(m_1) \hat f_i^\ast(k|m_1) \right] \
                \left[\int dm_2\,\frac{dn}{dm_2} b_j(m_2) \hat f_j(k|m_2) \right]\,P^{\rm lin}(k)\label{eq:PS2halo}
\eea
Notice that the average $\langle g \rangle$ of the density field of the source is given by:
\be
\bar g(z)=\langle g(z,\vec n) \rangle=\int dm\,\frac{dn}{dm}\int d^3\bm x \ g(\bm x|m,z)\;,
\label{eq:gave}
\ee
which implies that at small $k$ (where 
$\hat f\sim\int d^3\bm x\,g(\bm x|m)/\bar g $) the terms in the 
square-brackets in Eq.~(\ref{eq:PS2halo}) are of order 1 (except in the case of 
a significant bias). The 2-halo term is thus normalized to the standard linear 
matter PS at small $k$, which motivates the normalization of the window 
function introduced above.

In principle, instead of adopting the above halo-model formalism, one could directly take the results from N-body simulations, generate the corresponding maps of the emissions and then extract the angular power spectrum.
On the other hand, the small scales are currently not covered by simulations and in order to make a realistic prediction of the signal (in particular in the annihilating DM scenario) one has to introduce some prescription to include them. Results based on simulations only (without the introduction of small scales) can be significantly biased.
The halo-model approach is instead based on tuning three-functions (describing the number of halos $dn/dm$, the mass concentration of halos $c_{\rm vir}$, and the halo profile $\rho$) to the simulation results, where they exist, and defining some appropriate extrapolations in the unsampled regime.
The computation is therefore quite neat once those functions have been drawn, and allows for an easier physical insight.
Since it is probably unlikely that N-body simulations will cover scales corresponding to the WIMP free-streaming mass in the forthcoming years, such a phenomenological approach will remain a prime way to compute the WIMP angular PS. 

\section{Three-dimensional power spectra}

The next step is to explicitly derive the three-dimensional power spectra, defined in Eq.~(\ref{eq:PSs}), for the different cases of auto-correlation and cross-correlation involving particle DM signals. This requires to adopt, for each specific case, the appropriate intensity-field function $g$.

We will only consider a DM source with a spatial profile directly tracing the DM density profile (or its square in the annihilating scenario).
This is typically the case for prompt emission of gamma-rays, while for the radio emission, and for the X-ray or gamma-ray signal produced by inverse Compton processes, the radiative emissions associated to electrons and positrons injected by DM can have a different shape either because of the spatial profile of the fields of interaction (the magnetic field for the synchrotron emission which produces the radio signal; the radiation fields, like CMB or starlight, which are responsible for the X-ray, or also gamma-rays, production through inverse Compton scattering; both fields contribute to energy losses), or because of the diffusion of $e^+/e^-$ before emission. In this case the density field of the emission is no longer proportional to the DM density (or its square) and has to be accordingly modified.
This modification typically affects only anisotropies on very small scales (see the full formalism in Ref. \cite{Fornengo:2011xk}) and, for simplicity, will be neglected here except for some approximate estimate in the case of the radio emission.

\subsection{Auto-correlation}

\begin{figure}[t]
   \centering
   \includegraphics[width=0.49\textwidth]{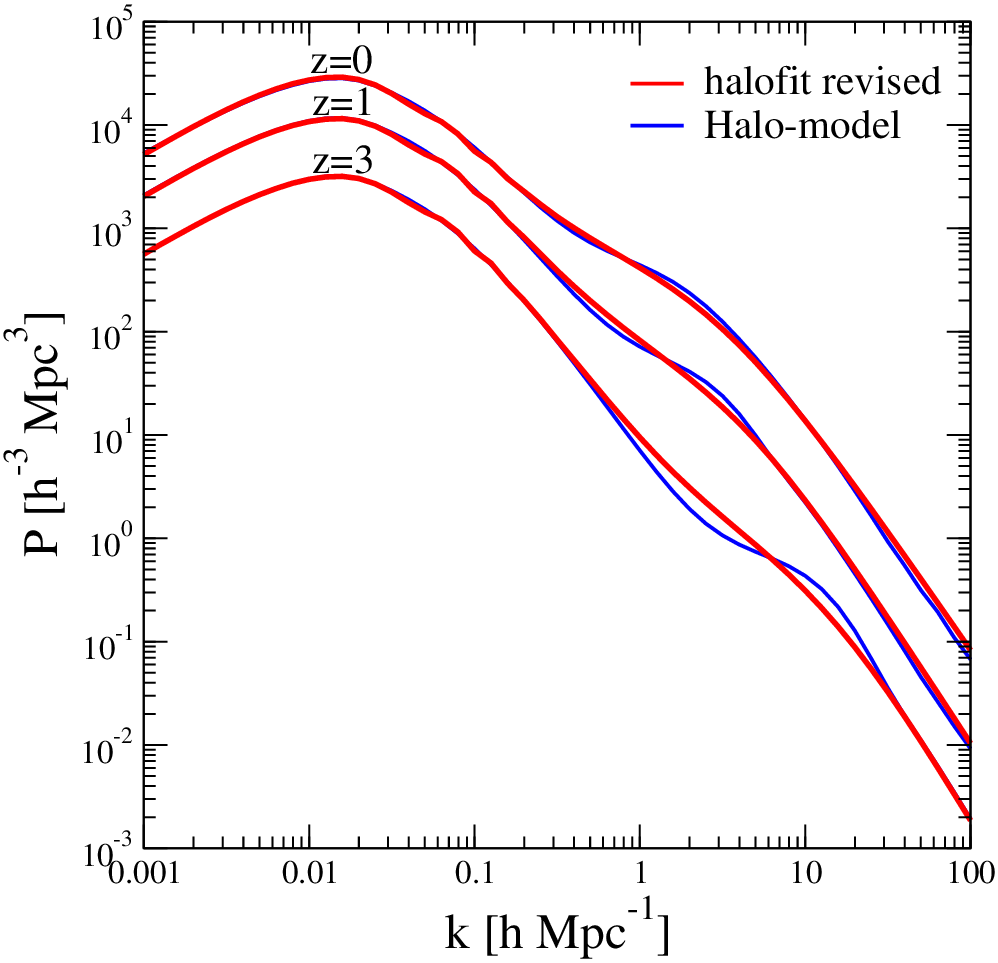}
\caption{{\bf 3D power spectra}. Comparison between the non-linear 3D matter-PS obtained with the halo-model considered in this paper and the revised halofit results from high-resolution $N$-body simulations presented in~\cite{Takahashi:2012em}, at different redshifts.}
\label{fig:ps1}
\end{figure}

In the cases of lensing and decaying DM,  the 
density field of the source is directly given by the density distribution $\rho(\bm x)$: $g(\bm x)=\rho(\bm x)$, and therefore
$f(\bm x)=\delta(\bm x)$, where the $\delta(\bm x)$ is the density contrast.
This is obvious for decaying DM, while in the lensing case it stems from the fact that gravitational lensing is due to the potential wells of the large 
scale structure, which are related to the matter distribution $\rho$ by the Poisson equation.
Therefore, from Eqs.~(\ref{eq:PS1halo}) and (\ref{eq:PS2halo}), the non-linear matter PS $P_{\delta\delta}$ is obtained by summing
up the following one-halo and two-halo terms:
\bea
 P_{\delta\delta}^{1h}(k) &=& \int_{m_{\rm min}}^{m_{\rm max}} dm\ \frac{dn}{dm} \tilde v(k|m)^2 \label{eq:psdelta1}
 \\
 P_{\delta\delta}^{2h}(k) &=& \left[\int_{m_{\rm min}}^{m_{\rm max}} dm\,\frac{dn}{dm} b_h(m) \tilde v(k|m) \right]^2\,P^{\rm lin}(k)
\label{eq:psdelta2}
\eea
where $\tilde v(k|m)$ is the Fourier transform of $\rho(\bm x|m)/\bar \rho$.
In the rest of the paper, we will adopt the halo mass function $dn/dm$ of Ref. \citep{Sheth:1999mn}, the halo
concentration from Ref. \citep{MunozCuartas:2010ig} down to $10^{10}M_\odot$ and extrapolating $c_{\rm vir}$ at smaller masses following Ref. \cite{Bullock:1999he}, and a NFW halo density 
profile~\citep{Navarro:1996gj}.
The halo bias $b_h$ is taken from Ref. \citep{Cooray:2002dia}.
We compare our halo-model-based predictions for the matter PS with latest results from high-resolution 
$N$-body simulations~\cite{Takahashi:2012em} in Fig.~\ref{fig:ps1}. 
A very good agreement is achieved at low redshift, which is the most relevant epoch for DM purposes, as we will see in the following.

For astrophysical sources (as, e.g., the LSS tracer considered below), which are better characterized by their luminosity $\mathcal{L}$ rather than the mass, the formalism described above can  still be adopted by simply replacing the mass function $dn/dm$ with the luminosity function $dn/d\mathcal{L}\equiv\Phi(\mathcal{L},z)$. 
Approximating astrophysical sources as point sources, we have 
$g_S(\mathcal{L},\bm x-\bm x')=\mathcal{L}\,\delta^3(\bm x-\bm x')$, and 
Eqs.~(\ref{eq:PS1halo}) and (\ref{eq:PS2halo})  give:
\bea
 P_{SS}^{1h}(k,z) &=& \int_{\mathcal{L}_{\rm min}(z)}^{\mathcal{L}_{\rm max}(z)} d\mathcal{L}\ \Phi(\mathcal{L},z)\,\left(\frac{\mathcal{L}}{\langle g_S \rangle}\right)^2
\\
 P_{SS}^{2h}(k,z) &=& \left[\int_{\mathcal{L}_{\rm min}(z)}^{\mathcal{L}_{\rm max}(z)} d\mathcal{L}\ \Phi(\mathcal{L},z)\, b_S(\mathcal{L},z)\,\frac{\mathcal{L}}{\langle g_S \rangle} \right]^2 \,P^{\rm lin}(k,z)\;.\label{eq:PSB}
\eea
Since the term in the square-brackets does not depend on $k$ (because of the 
point-source approximation) and Eq.~(\ref{eq:gave}) is now 
$\langle g_S \rangle= \int_{\mathcal{L}_{\rm min}(z)}^{\mathcal{L}_{\rm max}(z)} d\mathcal{L}\,\Phi\,\mathcal{L}$, the two-halo term 
is just a rescaled version of $P^{\rm lin}$ with the rescaling factor due to 
the bias (so possibly varying with $z$). The one-halo term is constant in $k$, 
namely it is a ``Poisson-noise" term.
The power spectrum of LSS tracers can be thus computed starting from the luminosity function of the population under investigation as described above (for gamma-ray emitters see, e.g., Ref. \cite{Camera:2012cj}).
However for a given survey might be not so easy to separate different populations (or impossible when dealing with unresolved contributions, unless using theoretical arguments).
To keep the discussion general, here we adopt a common description for all types of galaxies, using a halo occupation distribution of galaxies $N_{\rm gal}$ and weighting the contributions of different redshifts through the redshift distribution function $dn_{\rm gal}/dz$.
This means that the density field of galaxies can be expressed as $g({\bm x},M,z)=\rho({\bm x},M,z) \langle N_{\rm gal}(M,z)\rangle /\bar n_{\rm gal}(z)$, where $\bar n_{\rm gal}=\int dM\,dn/dM\, \langle N_{\rm gal}\rangle$ and $\rho$ is the host-halo density profile. The expression of $N_{\rm gal}$ is taken from Ref. \cite{Zheng:2004id}. In this approximated scenario, we have:
\bea
 P_{\rm gal,\rm gal}^{1h}(k) &=& \int_{m_{\rm min}}^{m_{\rm max}} dm\ \frac{dn}{dm} \frac{\langle N_{gal}\,(N_{gal}-1)\rangle}{\bar n_{gal}^2}\tilde v(k|m)^2 \label{eq:psLSS1}
 \\
 P_{\rm gal,\rm gal}^{2h}(k) &=& \left[\int_{m_{\rm min}}^{m_{\rm max}} dm\,\frac{dn}{dm} b_h(m) \frac{\langle N_{gal}\rangle}{\bar n_{gal}} \tilde v(k|m) \right]^2\,P^{\rm lin}(k)\;.
\label{eq:psLSS2}
\eea

In the case of annihilating DM, the signal scales with the square of the density field:
$g(\bm x)=\rho^2(\bm x)$. In order to derive the 
associated angular PS, we can use the formalism outlined in the 
previous section where now we square the expression given in Eq.~(\ref{eq:denseed}) (also
here $f(\bm x)=\rho(\bm x)$, as in the decaying DM case). The correlation function would then lead to a term proportional to:
\be
\langle \sum_i\, \delta^3(\bm x_1-\bm x_i)\delta(m_1-m_i)\,\sum_j\, \delta^3(\bm x_2-\bm x_j)\delta(m_2-m_j)\sum_k\, \delta^3(\bm x_3-\bm x_k)\delta(m_3-m_k)\,\sum_l\, \delta^3(\bm x_4-\bm x_l)\delta(m_4-m_l) \rangle
\ee
which in turn can be expanded into four-, three-, and two-point seed 
correlation functions.
However, since DM halos (described by $\rho(\bm x|m)$) are mutually exclusive, 
an annihilation can only occur if the two particles are within the same halo. 
This implies that all the terms other than the one- and two-halo contributions 
(which are the same appearing in Eq.~(\ref{eq:2corr})) vanish. Therefore,
Eq.~(\ref{csi2x}) does not formally change, with $f_a$ now being 
$\rho(\bm x|m)^2/\langle \rho^2 \rangle$.
The auto-correlation PS $P_{\delta^2\delta^2}$ is thus given by the sum of the two
following terms:
\bea
P_{\delta^2\delta^2}^{1h}(k) &=& \int_{m_{\rm min}}^{m_{\rm max}} dm\ \frac{dn}{dm} \left(\frac{\tilde u(k|m)}{\Delta^2}\right)^2 \label{eq:PSh1}
\\
P_{\delta^2\delta^2}^{2h}(k) &=& \left[\int_{m_{\rm min}}^{m_{\rm max}} dm\,\frac{dn}{dm} b_h(m) \frac{\tilde u(k|m)}{\Delta^2} \right]^2\,P^{\rm lin}(k)
\label{eq:PSh2}
\eea
where $\tilde u(k|m)$ is the Fourier transform of 
$\rho^2(\bm x|m)/\bar \rho^2$.
Our results are only very mildly dependent on the upper mass cutoff (which 
we set to be $m_{\rm max}=10^{18}M_{\odot}$) and the clustering on large scales has been well sampled by means of simulations and observations.
A crucial and uncertain ingredients is instead the DM clustering at very small masses
(minimum halo mass $m_{\rm min}$, concentration parameter, and substructure scheme).
Concerning substructures, we include them by adopting the scheme described in Ref. \cite{Kamionkowski:2010mi}, namely by replacing $\rho^2(\bm x,m,z)$ with $B(\bm x,m,z)\rho^2(\bm x,m,z)$, where $B(\bm x,m,z)$ is a boost function associated to subhalos which multiplies the smooth halo density profile.

\begin{figure}[t]
   \centering
   \includegraphics[width=0.45\textwidth]{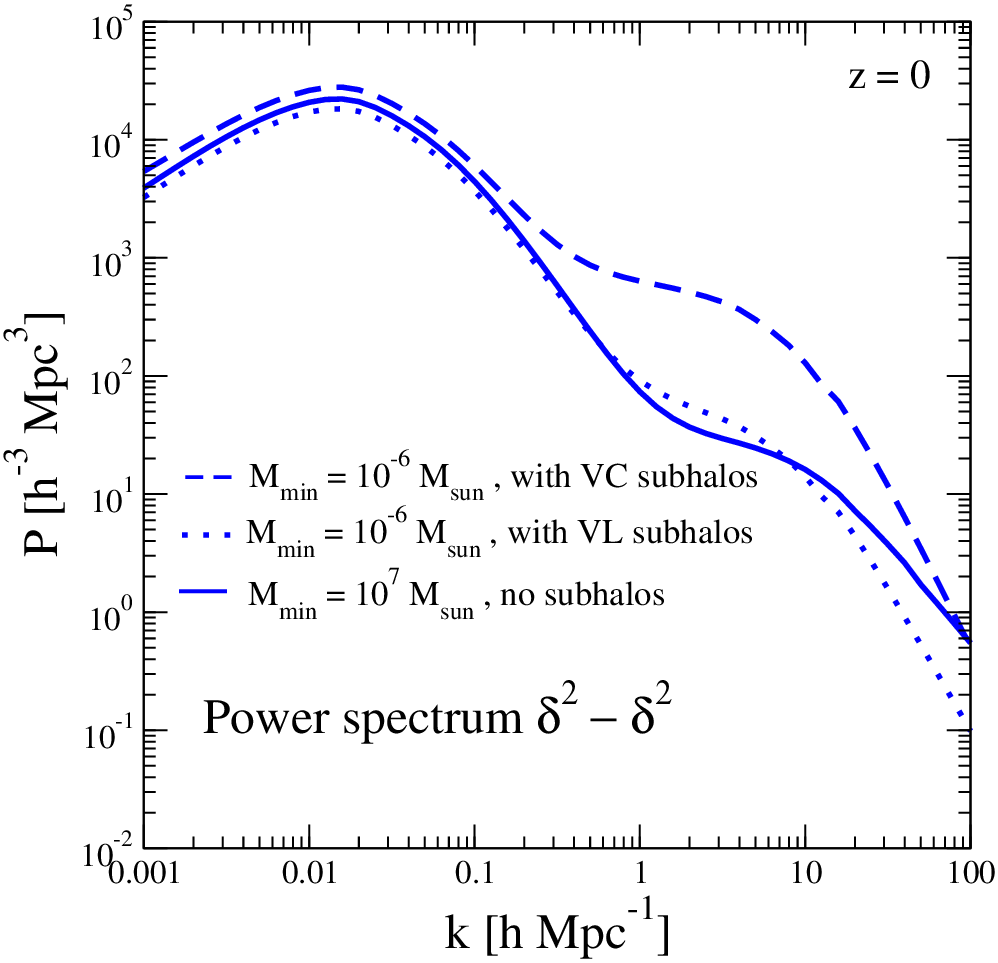}\hspace{0.5cm}
   \includegraphics[width=0.45\textwidth]{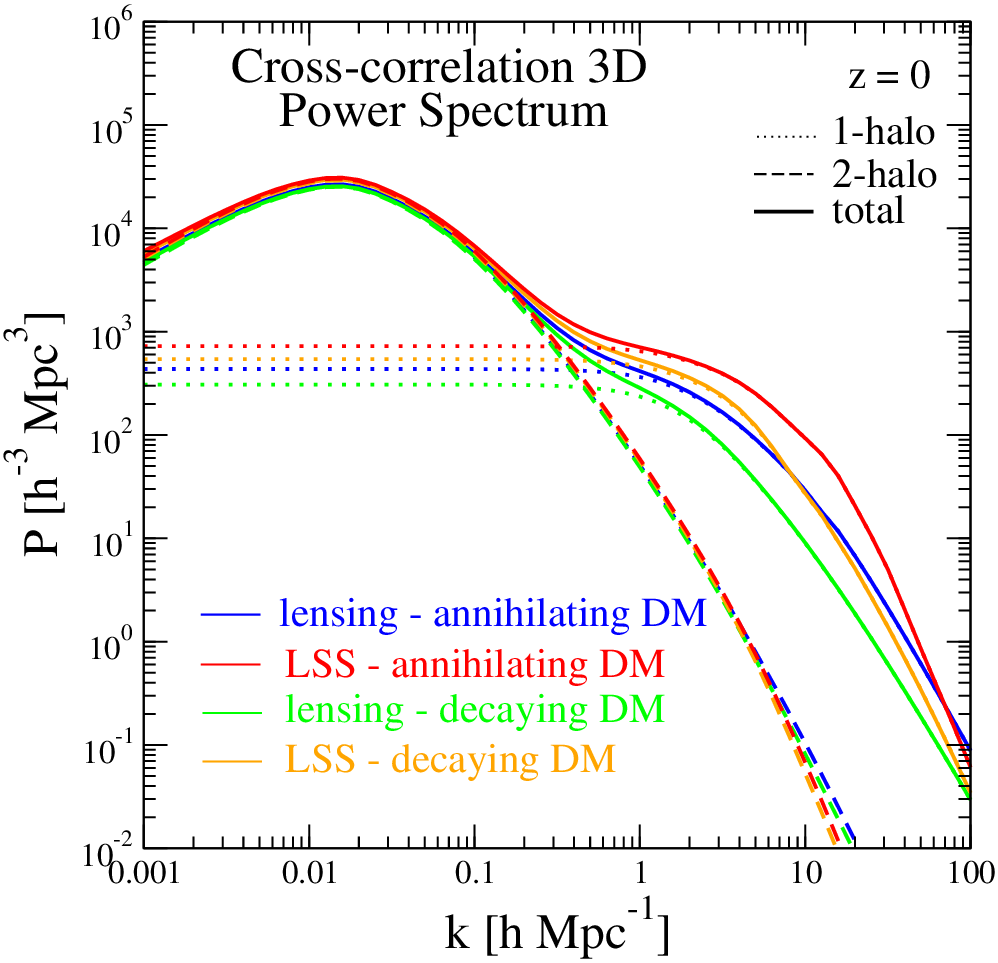}
\caption{{\bf 3D power spectra}.  
{\sl Left:} Auto-correlation 3D PS of annihilating DM in the three scenarios described in the text, at $z=0$. {\sl Right:} Cross-correlation 3D PS of annihilating/decaying DM with lensing and galaxy tracers; the total, 1-halo and 2-halo terms are explicitly shown.}
\label{fig:ps2}
\end{figure}

In the left panel of Fig.~\ref{fig:ps2}, we show three specific examples of 3D power spectra, which refer
to DM models endowed with different features: two models with 
$m_{\rm min}=10^{-6} M_{\odot}$ (which is the typical WIMP free-streaming mass) 
with substructures described by adapting the boost-function $B$ to reproduce the Via Lactea (VL) simulation~\cite{Kamionkowski:2010mi} or the Virgo Collaboration (VC) results~\cite{Fornasa:2012gu}); one model with $m_{\rm min}=10^{7}M_{\odot}$ (which is the minimum halo mass currently inferred from dynamical measurements) without substructures.
Having normalized the power spectra to be (roughly) equal on large scales, the effect of substructures mainly reflects into boosting the power on Mpc-scales.
In the VC scenario the signal from largest halos is indeed strongly boosted by the contribution of subhalos, while in the VL scenario the effect is much milder.

\subsection{Cross-correlation}

In the case of cross-correlations of electromagnetic DM signals
among themselves, the situation is similar to the one discussed in the previous Section for auto-correlations: for decaying DM, since all signals depend linearly on the DM density, the relevant matter PS is $P_{\delta\delta}$, whose one- and two-halo terms are given by Eqs. (\ref{eq:psdelta1},\ref{eq:psdelta2}), while for annihilating DM the matter PS is 
$P_{\delta^2\delta^2}$, with its component given by Eqs. (\ref{eq:PSh1},\ref{eq:PSh2}).

In the case of cross-correlation between lensing (which depends linearly on the DM density) and an electromagnetic signal produced by a decaying DM particle, we are again
concerned with the $P_{\delta\delta}$ power spectrum, expressed through Eqs. (\ref{eq:psdelta1},\ref{eq:psdelta2}). 

The case of cross-correlation of astrophysical sources 
with decaying DM is analogous to the case involving astrophysical 
sources and lensing signal, both given by:
\bea
 P_{S\delta}^{1h}(k,z) &=& \int_{\mathcal{L}_{\rm min}(z)}^{\mathcal{L}_{\rm max}(z)} d\mathcal{L}\,\Phi(\mathcal{L},z)\,\frac{\mathcal{L}}{\langle g_S \rangle} \,\tilde v(k|m(\mathcal{L}))  \\
 P_{S\delta}^{2h}(k,z) &=& \left[\int_{\mathcal{L}_{\rm min}(z)}^{\mathcal{L}_{\rm max}(z)} d\mathcal{L}\,\Phi(\mathcal{L},z)\, b_S(\mathcal{L},z)\,\frac{\mathcal{L}}{\langle g_S \rangle} \right]\left[\int_{m_{\rm min}}^{m_{\rm max}} dm\,\frac{dn}{dm} b_h(m)\,\tilde v(k|m) \right] \,P^{\rm lin}(k,z)\;.\label{eq:PSBd2}
\eea
Notice that this requires a relation between the source luminosity $\mathcal{L}$ and  the host-halo mass $m$.
In the approximate description introduced above, we have instead:
\bea
 P_{\rm gal,\delta}^{1h}(k) &=& \int_{m_{\rm min}}^{m_{\rm max}} dm\ \frac{dn}{dm} \frac{\langle N_{gal}\,\rangle}{\bar n_{gal}}\tilde v(k|m)^2 \label{eq:psLSSdelta1}
 \\
 P_{\rm gal,\delta}^{2h}(k) &=& \left[\int_{m_{\rm min}}^{m_{\rm max}} dm\,\frac{dn}{dm} b_h(m) \frac{\langle N_{gal}\rangle}{\bar n_{gal}} \tilde v(k|m) \right] \left[\int_{m_{\rm min}}^{m_{\rm max}} dm\,\frac{dn}{dm} b_h(m)\,\tilde v(k|m) \right]\,P^{\rm lin}(k)\;.
\label{eq:psLSSdeta2}
\eea

In the case of cross-correlation between a gravitational tracer and a signal from DM annihilation, the matter power spectrum $P_{\delta\delta^2}$ involves terms of the type:
\be
\langle \sum_i\, \delta^3(\bm x_1-\bm x_i)\delta(m_1-m_i)\,\sum_j\, \delta^3(\bm x_2-\bm x_j)\delta(m_2-m_j)\sum_k\, \delta^3(\bm x_3-\bm x_k)\delta(m_3-m_k) \rangle\,,
\ee
For the same reasons discussed above in connection with the auto-correlation of
annihilating DM, only the terms in the r.h.s. of Eq.~(\ref{eq:2corr}) survive,
leading again to Eq.~(\ref{csi2x}) for the two-point correlation function
(see also Appendix B in Ref. \cite{Ando:2006cr}). This allows to write the one- and
two-halo terms of the PS  of cross-correlation with lensing or decaying DM as:
\bea
P_{\delta\delta^2}^{1h}(k) &=& \int_{m_{\rm min}}^{m_{\rm max}} dm\ \frac{dn}{dm} \tilde v(k|m) \,\frac{\tilde u(k|m)}{\Delta^2} \label{eq:PSsh1} \\
P_{\delta\delta^2}^{2h}(k) &=& \left[\int_{m_{\rm min}}^{m_{\rm max}} dm\,\frac{dn}{dm}b_h(m) \tilde v(k|m) \right]\,\left[\int_{m_{\rm min}}^{m_{\rm max}} dm\,\frac{dn}{dm} b_h(m) \frac{\tilde u(k|m)}{\Delta^2} \right]\,P^{\rm lin}(k)\;,
\label{eq:PSsh2}
\eea
The PS in the case of cross-correlation of annihilating DM with astrophysical sources is:
\bea
 P_{S\delta^2}^{1h}(k,z) &=& \int_{\mathcal{L}_{\rm min}(z)}^{\mathcal{L}_{\rm max}(z)} d\mathcal{L}\,\Phi(\mathcal{L},z)\,\frac{\mathcal{L}}{\langle g_S \rangle} \,\frac{\tilde u(k|m(\mathcal{L}))}{\Delta^2} \nonumber\\
 P_{S\delta^2}^{2h}(k,z) &=& \left[\int_{\mathcal{L}_{\rm min}(z)}^{\mathcal{L}_{\rm max}(z)} d\mathcal{L}\,\Phi(\mathcal{L},z)\, b_S(\mathcal{L},z)\,\frac{\mathcal{L}}{\langle g_S \rangle} \right]\left[\int_{m_{\rm min}}^{m_{\rm max}} dm\,\frac{dn}{dm} b_h(m) \frac{\tilde u(k|m)}{\Delta^2} \right] \,P^{\rm lin}(k,z)\;,
\label{eq:PSBh}
\eea
and in the approximate scenario is:
\bea
P_{\rm gal,\delta^2}^{1h}(k) &=& \int_{m_{\rm min}}^{m_{\rm max}} dm\ \frac{dn}{dm} \frac{\langle N_{gal}\rangle}{\bar n_{gal}} \tilde v(k|m) \,\frac{\tilde u(k|m)}{\Delta^2} \label{eq:PSannLSS1} \\
P_{\rm gal,\delta^2}^{2h}(k) &=& \left[\int_{m_{\rm min}}^{m_{\rm max}} dm\,\frac{dn}{dm}b_h(m) \frac{\langle N_{gal}\rangle}{\bar n_{gal}} \tilde v(k|m) \right]\,\left[\int_{m_{\rm min}}^{m_{\rm max}} dm\,\frac{dn}{dm} b_h(m) \frac{\tilde u(k|m)}{\Delta^2} \right]\,P^{\rm lin}(k)\;,
\label{eq:PSannLSS2}
\eea

A few examples of power spectra for the different cross-correlation cases are shown in Fig.~\ref{fig:ps2}, for both annihilating and decaying DM with lensing or galaxy tracers.

\begin{figure}[t]
\centering
\includegraphics[width=0.45\textwidth]{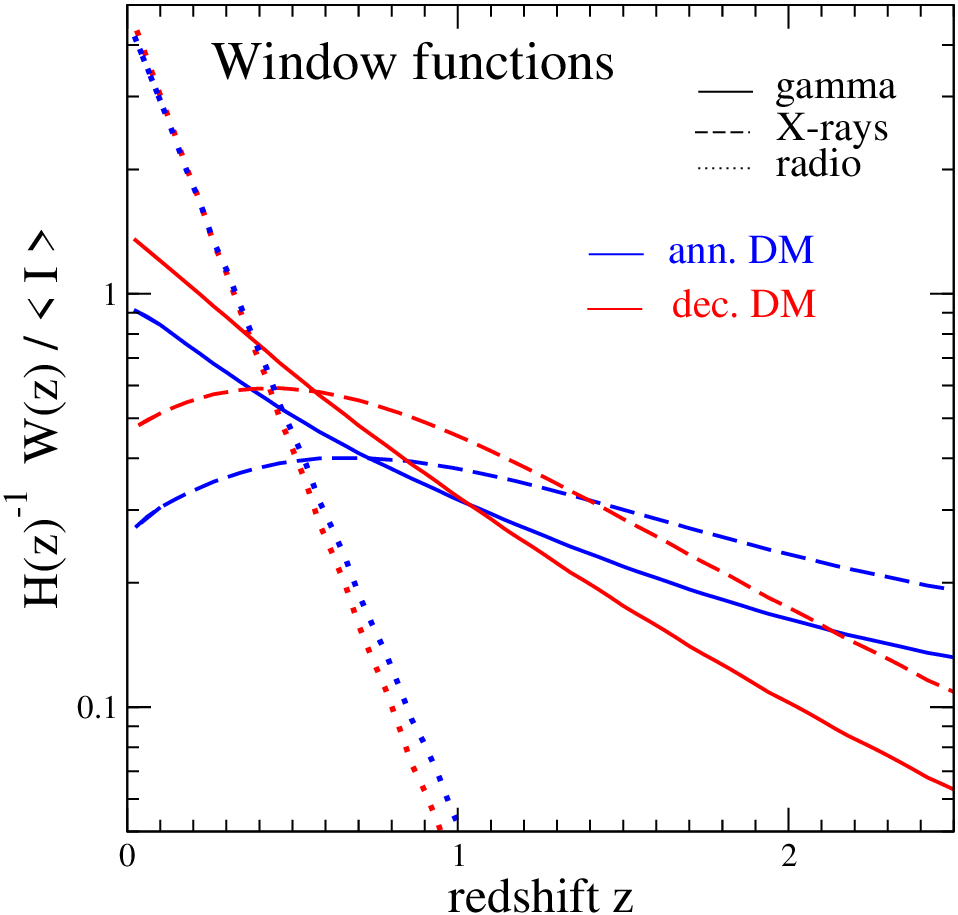}\hspace{0.5cm}
\includegraphics[width=0.45\textwidth]{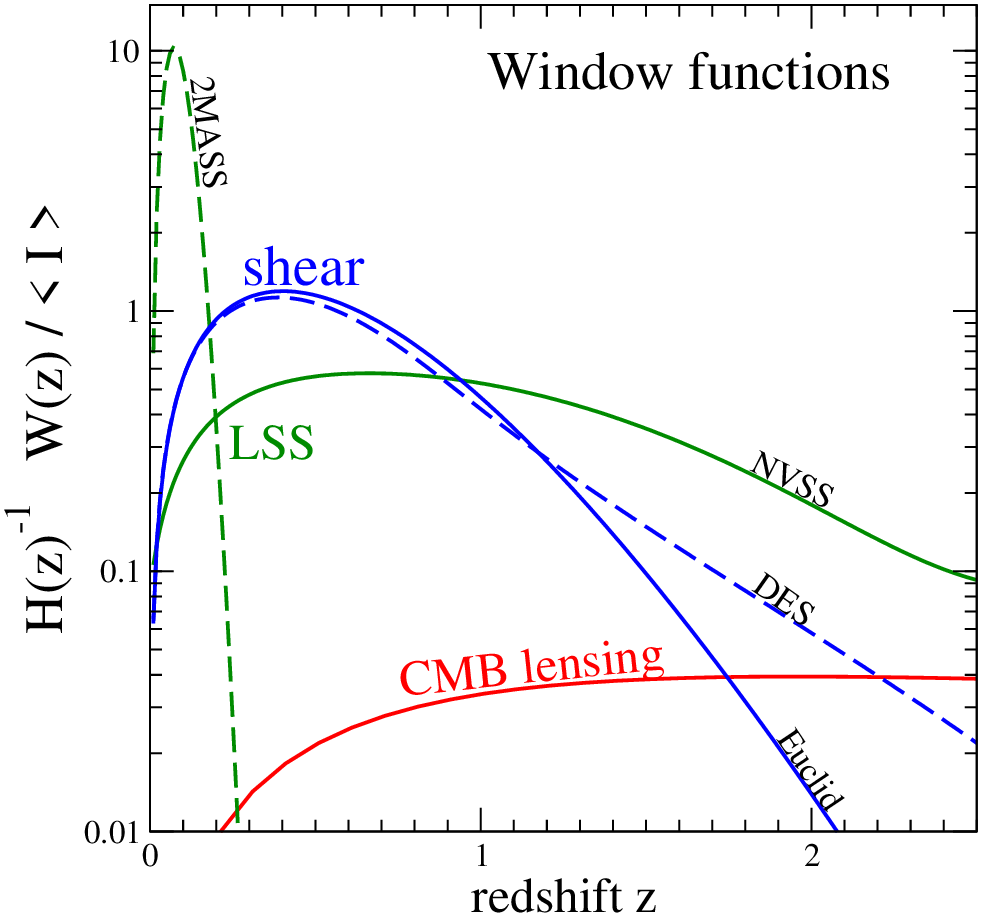}
\caption{{\bf Window functions}. {\sl Left}: Window functions as a function of the redshift for radio, X-ray, and gamma-ray emissions from annihilating/decaying DM. The window functions
refer to reference frequencies/energies: $\nu=1$ GHz for radio, $E=10$ keV for X-rays, $E=1$ GeV for gamma-rays.
They have been normalized to the total average intensity $\langle I \rangle$ such that the integral of the curves is equal to 1 for all cases. {\sl Right:}  Window functions vs. redshift for cosmic shear, CMB lensing, and two populations of galaxies (detected in the 2MASS and NVSS surveys) tracing LSS.
}
\label{fig:winfun}
\end{figure}

\section{Window functions}
\label{sec:window}

The final step is to determine the window functions of the different observables. Let us start with the gravitational tracers. The window function of the lensing signal takes the form (see, e.g., Ref.~\citep{Bartelmann:2010fz}):
\be
W(\chi)=\frac{3}{2}\ho^2\om[1+z(\chi)]\chi\int_\chi^\infty\!\!\de\chi'\,\frac{\chi'-\chi}{\chi'}\frac{\de N}{\de\chi'}(\chi')
\label{eq:W(z)}
\ee
where $H_0$ is the Hubble constant, $\om$ the matter-density parameter and $\de N/\de \chi$ denotes the redshift distribution of the background sources (which are lensed by the structures), normalized to unity area.

In the case of CMB lensing, the source is provided by the last scattering surface. Approximating it as a surface with infinitesimal width located at $z_*$, we can use the same Eq.~(\ref{eq:W(z)}) with $\de N/\de z=\delta(z-z_*)$. In this case we obtain:
\be
W(\chi)=\frac{3}{2}\ho^2\om[1+z(\chi)]\chi\,\frac{\chi_*-\chi}{\chi_*}
\label{eq:wcmb}
\ee

In the case of astrophysical sources, as e.g. the LSS tracers, we have  (see, e.g., Ref.~\cite{Camera:2012cj}):
\be
W(E,z) =  \frac{\langle g_S(z)\rangle}{4\,\pi\,(1+z)} \, e^{-\tau[E(1+z),z]} \;.
\label{eq:wastro}
\ee
However, in this paper, following the discussion outlined in the previous section, we employ a common description for all types of galaxies with a window function given by $W= dn_{\rm gal}/dz$.

For the case of decaying DM we again have 
$f(\bm x)=\delta(\bm x)$ and the window 
function is given by (see, e.g., Ref.~\cite{Ibarra:2009nw}):
\be
W(E,z) = \frac{1}{4\pi} \frac{\Omega_{\rm DM}\rho_c}{m_\chi \tau_d} \frac{dN_d[E(1+z)]}{dE} e^{-\tau[E(1+z),z]}
\label{eq:wdec}
\ee
where $\Omega_{\rm DM}$ is the DM density parameter, $\rho_c$ is the critical density
of the Universe, $m_\chi$ and $\tau_d$ denote the mass and decay lifetime
of the DM particle, respectively, and $dN_d[E]/dE$ is the number of photons 
(at radio, X-ray, or $\gamma$-ray frequency, depending in the specific signal under study) emitted per decay-event in the energy range $(E,E+dE)$. Finally, $\tau$ is the optical depth for 
absorption.
It can be relevant in the gamma-ray case and is mainly due to pair production on
the extragalactic background light emitted by galaxies in the ultraviolet, optical, and infrared bands.
For radio and X-rays, the absorption can have an impact only in the very low frequency part of the bands.
Notice that the factor $\Omega_{\rm DM}\rho_c$ comes from the normalization of $W$, 
such that $\langle g_d \rangle=\bar \rho$. Notice also that we are considering 
the differential energy flux as the intensity $I$ under study: thus we quote differential (in energy) window functions for the DM electromagnetic signals.

In the case of annihilating DM, the signal scales with $\rho^2$, and to define the corresponding window function we can make use of the so-called clumping factor $\Delta^2(z)$, defined as:
\be
\Delta^2(z)=\frac{\langle \rho^2 \rangle}{\bar \rho^2}=\int_{m_{\rm min}}^{m_{\rm max}} dm\,\frac{dn}{dm}\int\ d^3\bm x\,\frac{\rho^2(\bm x|m)}{\bar \rho^2}\;,
\label{eq:clumpfact}
\ee
The window function for annihilating DM then takes the form  (see, e.g., Ref.~\cite{Ando:2005xg}):
\be
W(E,z) = \frac{(\Omega_{DM}\rho_c)^2}{4\pi} \frac{\langle\sigma_a v\rangle}{2 m_\chi^2} (1+z)^3\,\Delta^2(z)\,\frac{dN_a[E(1+z)]}{dE} \,e^{-\tau[E(1+z),z]}
\;,
\label{eq:wann}
\ee
where $\langle\sigma_a v\rangle$ is the velocity-averaged annihilation cross sections
times relative velocity (which we assume to be the same in all DM structures) and 
$dN_a[E]/dE$ is the number of photons emitted per annihilation event
in the energy range $(E,E+dE)$.

\begin{figure}[t]
\centering
\includegraphics[width=0.45\textwidth]{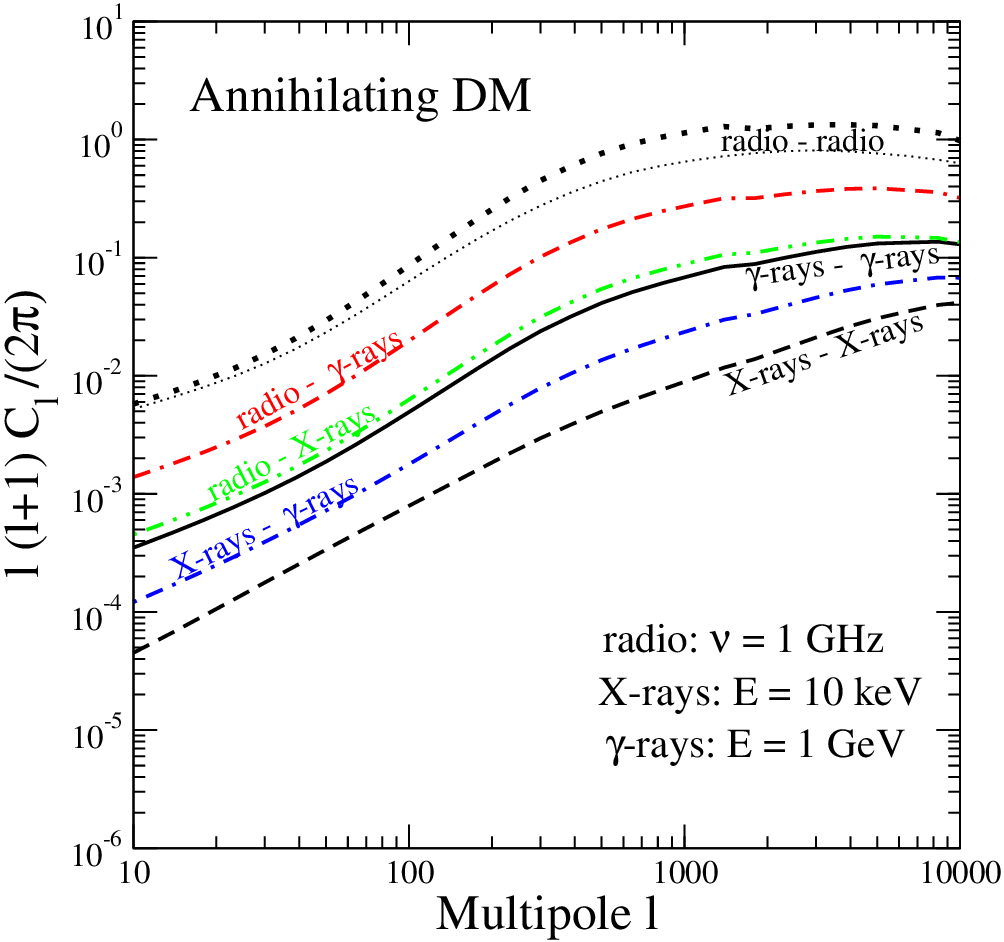}\hspace{0.5cm}
\includegraphics[width=0.45\textwidth]{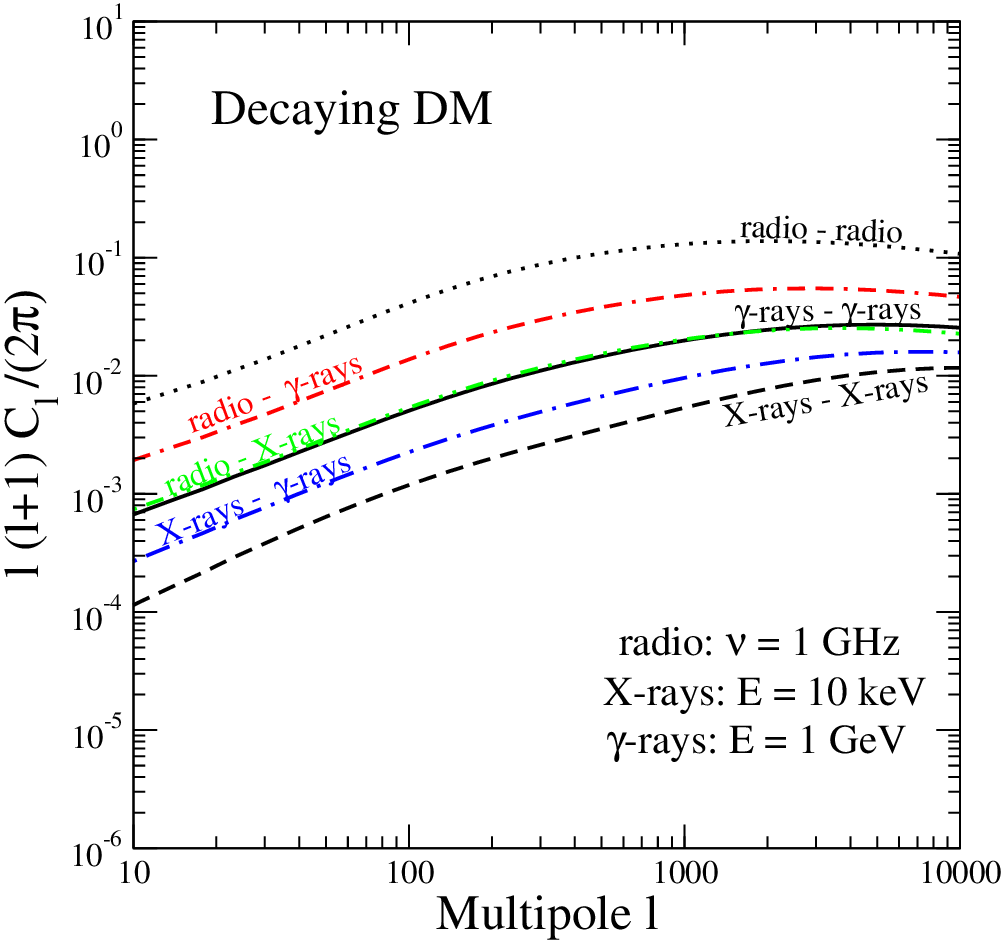}
\caption{Angular PS of radio, X-ray, and gamma-ray emissions (auto- and cross-correlations) from annihilating DM (left panel) and decaying DM (right panel). The computation is performed using the 3D PS models reported in Fig.~\ref{fig:ps1} and the window functions shown in Fig.~\ref{fig:winfun}. For the radio-radio case, we also show (with the thinner dotted line) the effect of a cored DM profile replacing the NFW distribution (as a possible results of particle diffusion).
}
\label{fig:corr1}
\end{figure}

\section{Discussion}

Let us turn now to the discussion of some examples of auto- and cross-correlation
angular power spectra. To be definite, and for illustrative purposes, we focus on a benchmark DM candidate with mass $m_{\chi}=100$ GeV, and final state of annihilation/decay into quarks $b-\bar b$.
Since we will report dimensionless window functions and angular power spectra
(i.e. quantities relative to the total average intensity),
the actual values of the annihilation cross section or the decay lifetime are not relevant in the discussion. For definiteness,
the substructure description is chosen to follow the VC scenario mentioned above.

Concerning LSS tracers, we will consider two specific examples,
given by the population of galaxies observed by the 2MASS~\cite{Skrutskie:2006wh} and NVSS~\cite{Condon:1998iy} surveys. 
The parameters entering in the window function $dn_{\rm gal}/dz$ of the two cases are taken from Ref. \cite{Afshordi:2003xu} and Ref. \cite{DeZotti:2009an}. 

In the case of weak-lensing shear, we consider the distributions of background sources that will be relevant for the Dark Energy Survey (DES) telescope and for the Euclid satellite.
The source distribution for DES \cite{Abbott:2005bi} can be described by $dN/\de z=A_D\,(z^a+z^{ab})/(z^b+c)$, with $a$, $b$, and $c$ provided in Table 1 of Ref. \citep{Fu:2007qq}, and $A_D$ fixed by the normalization condition $\int dz\, \de N/\de z=1$. For Euclid \cite{Euclid}, we can use $dN/\de z=A_E\,z^2\,\exp[-(z/z_0)^{1.5}]$, where $z_0=z_m/1.4$ with 
$z_m=0.9$ being the median redshift of the survey and $A_E$ again fixed by the 
normalization.

In the right panel of Fig.~\ref{fig:ps2} we show the cross-correlation 3D PS of annihilating/decaying DM with lensing and LSS tracers, computed as described in the previous Sections. 
Notice that the PS for the cross-correlation between lensing with decaying DM is the customary matter power spectrum $P_{\delta\delta}$ (described in Eq.~(\ref{eq:psdelta1},\ref{eq:psdelta2})).
The other cases (LSS with decaying DM and the cases with annihilating DM) lie above it, having more power on the structure scales $k\gtrsim 1\, {\rm Mpc}^{-1}$.
This is particularly true in the annihilating case and can be also understood by comparing Fig.~\ref{fig:ps1} with the left panel of Fig.~\ref{fig:ps2} in the VC scheme. 

The window functions are shown in Fig.~\ref{fig:winfun}.
The left panel refers to the particle DM cases and shows the redshift dependence of the window functions for the radio, X-ray, and gamma-ray emissions, in both annihilating and decaying scenarios. 
The radio signal is due to synchrotron-emission produced by electrons and positrons (induced by the DM annihilations/decays) interacting with a magnetic field chosen to be constant and with strength $B=10\,\mu G$. The X-ray flux instead arises from inverse Compton (IC) of the same electrons and positrons on the CMB, with ensuing up-scatter of the CMB photons. The window
functions depend on the photon energy, as is clear from Eqs. (\ref{eq:wdec},\ref{eq:wann}): for
definiteness, we choose some reference frequency/energy, such that the different emissions are all involving GeV particles (photons in the gamma-ray case, while $e^+/e^-$ in the radio and X-ray case) originated from the annihilations/decays. The reference frequencies are: $\nu=1$ GHz for radio, $E=10$ keV for X-rays, $E=1$ GeV for gamma-rays.
The plot shows that a multi-wavelength approach can be quite powerful. Indeed, the different emissions exhibit rather different window functions: each one therefore can provide different and complementary information.
In particular, the radio case is more peaked at low redshift, while the X-ray window function has a flatter shape, with the gamma-ray case being somewhat in between.
This is due to the fact that the CMB becomes much more dense in the past (the photon energy-density goes with the well known $(1+z)^4$ scaling), making the energy losses for IC very effective at higher redshifts: this depletes the radio signal at high $z$, while, in the X-ray case, the corresponding increase of the IC emissivity compensates the effect.
We also notice that the decaying and annihilating DM cases produce very similar window functions for radio emission, while for X-rays and gamma-rays the window function is flatter in the annihilating case as compared to the decaying case. This
occurs because of the effect of clustering, which enhances the signal at larger  redshifts.

The window functions for weak-lensing shear (DES and EUCLID populations), CMB lensing, and LSS (2MASS and NVSS galaxy catalogs) are shown in the right panel of Fig.~\ref{fig:winfun}. They have 
quite different shapes, as a consequence of the different redshift evolution of the relevant tracers which
produce the observed signal.
Since the cross-correlation power spectrum depends on the overlap of the window functions of the
two correlated signals, as is shown in Eq. (\ref{eq:clfin}), the different gravitational tracers
will be more/less efficient in the cross-correlation with particle DM signals. Comparing the two
panels of Fig. ~\ref{fig:winfun} allows to understand which cross-correlation
combinations could be optimal (although experimental limitations may occur, like limited statistics in one observational channel or angular resolutions).
Notice that 2MASS and NVSS are sort of two opposite limits for galaxy tracers (and this is the why we choose these examples):  2MASS coverage peaks at very low-redshifts, below 0.2, while NVSS
has a broad redshift distribution. The weak-lensing shear is peaked at intermediate-low redshift, which makes this observables a good candidate for cross-correlation with particle DM signals, as proposed  in~Ref. \cite{Camera:2012cj}. The CMB window function instead extends over a very wide range of redshift (in principle, up to the last scattering): this is also why it appears significantly lower, as compared to the other cases in Fig. \ref{fig:winfun}, since we have normalized all window functions to the total average intensity (i.e. the curves shown in Fig. \ref{fig:winfun} are normalized to 1).
In the following, we also normalize all the angular PS to the total averaged intensity, as reported in 
Eq.~(\ref{eq:clfin}), in order to make the various cases more easily comparable. The results will therefore
show the fractional size of the anisotropy signal (multiplied by $\ell(\ell+1)/(2\,\pi)$) with respect to the average isotropic intensity field.

\begin{figure}[t]
\centering
\includegraphics[width=0.45\textwidth]{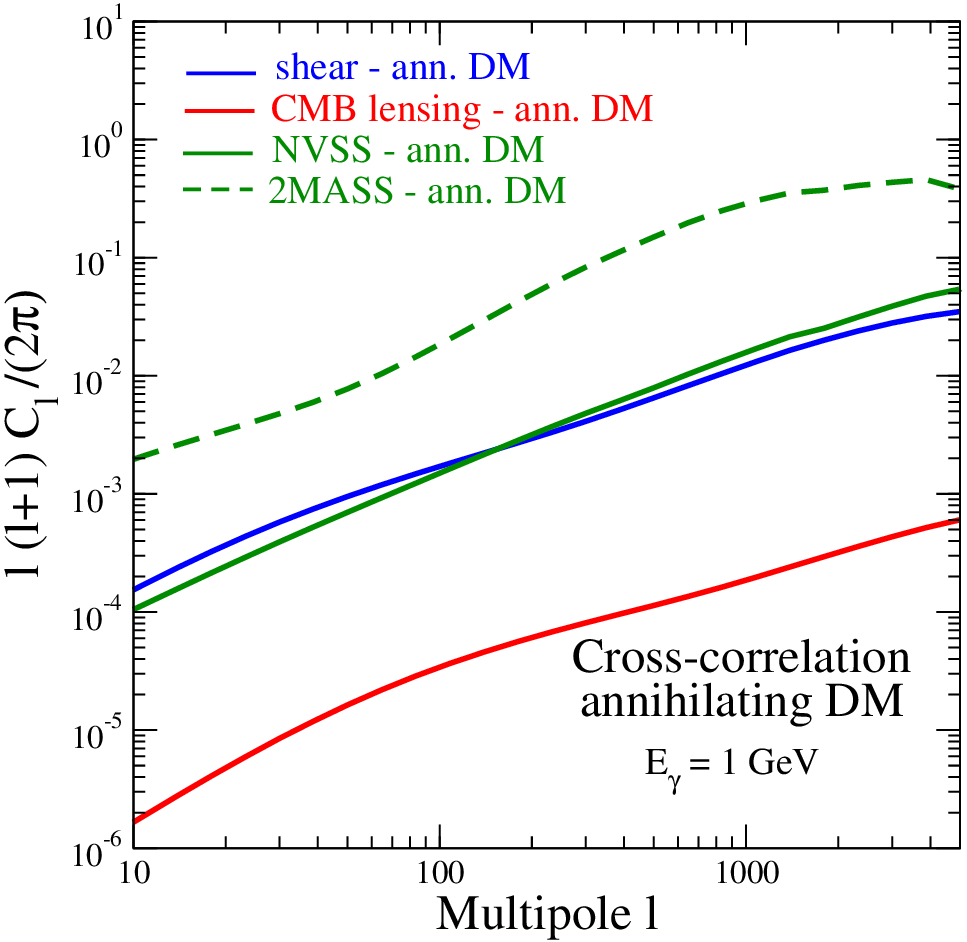}\hspace{0.5cm}
\includegraphics[width=0.45\textwidth]{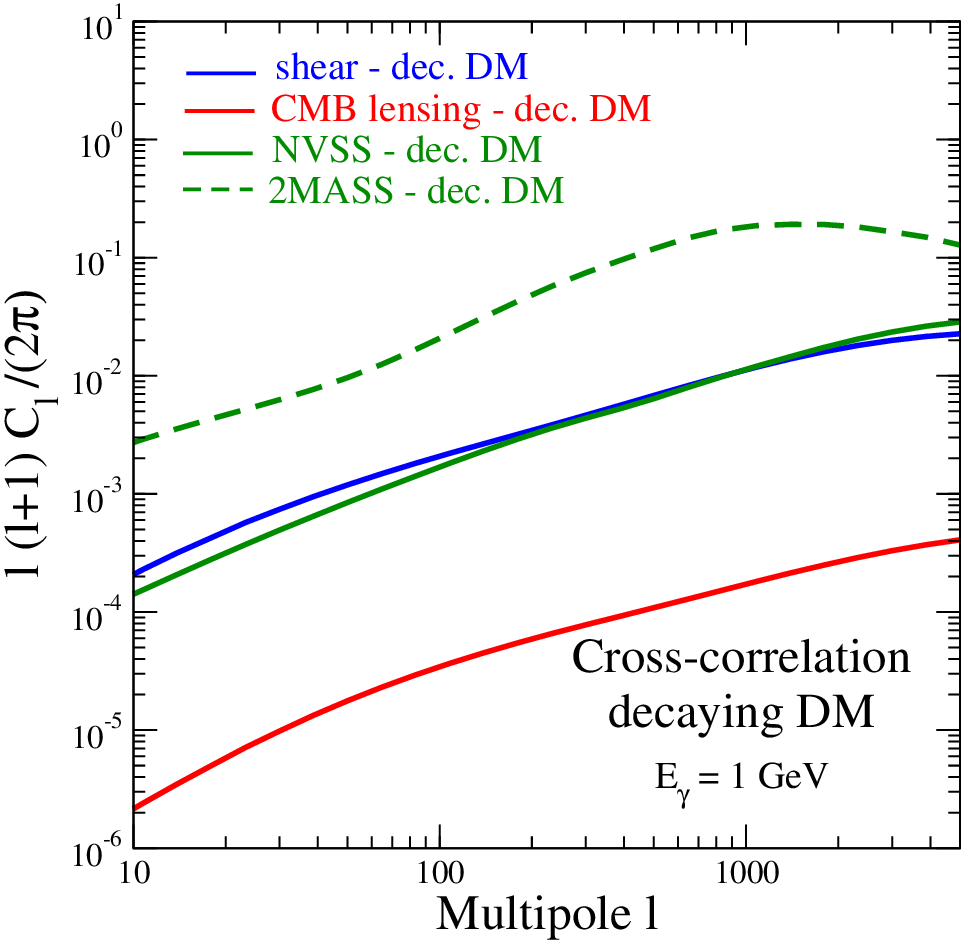}
\caption{Angular cross-correlation PS between gamma-rays (at $E_\gamma = 1$ GeV) and gravitational tracers (cosmic shear, CMB lensing and LSS tracers), for annihilating DM (left panel) and decaying DM (right panel). The computation is performed using the 3D PS models reported in Fig.~\ref{fig:ps2} and the window functions shown in Fig.~\ref{fig:winfun}. 
}
\label{fig:corr2}
\end{figure}

\begin{figure}[t]
\centering
\includegraphics[width=0.45\textwidth]{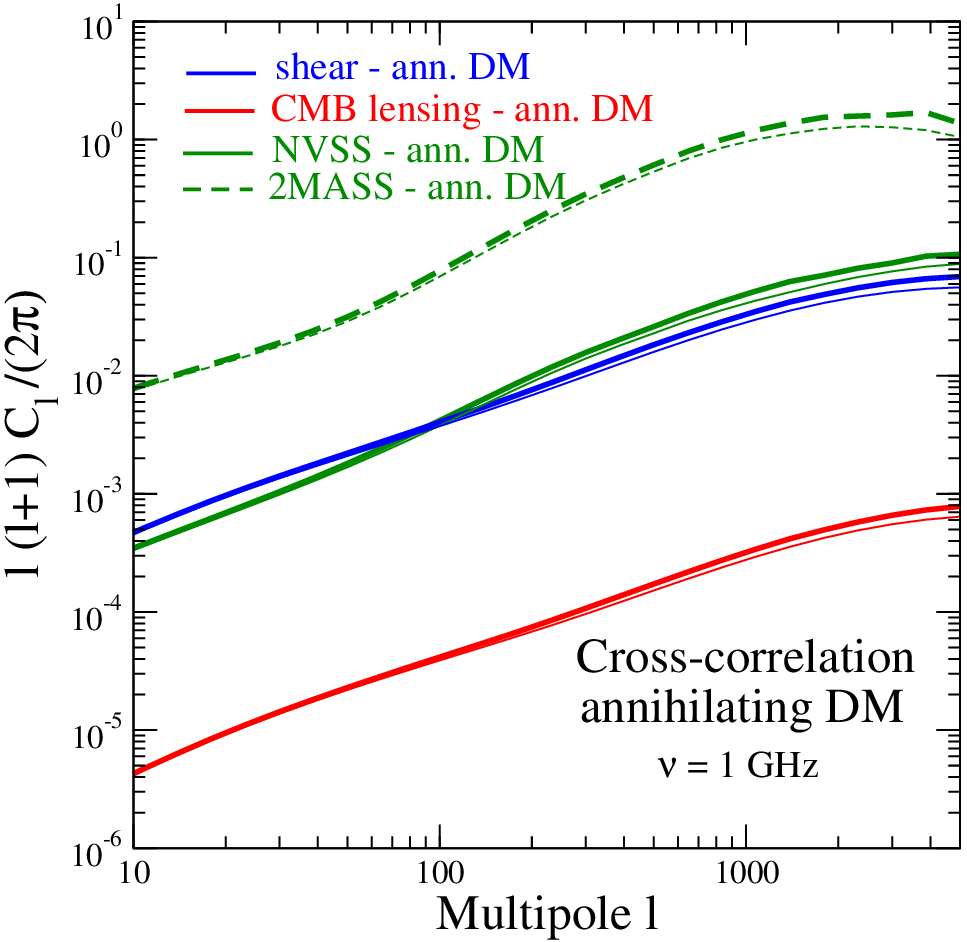}\hspace{0.5cm}
\includegraphics[width=0.45\textwidth]{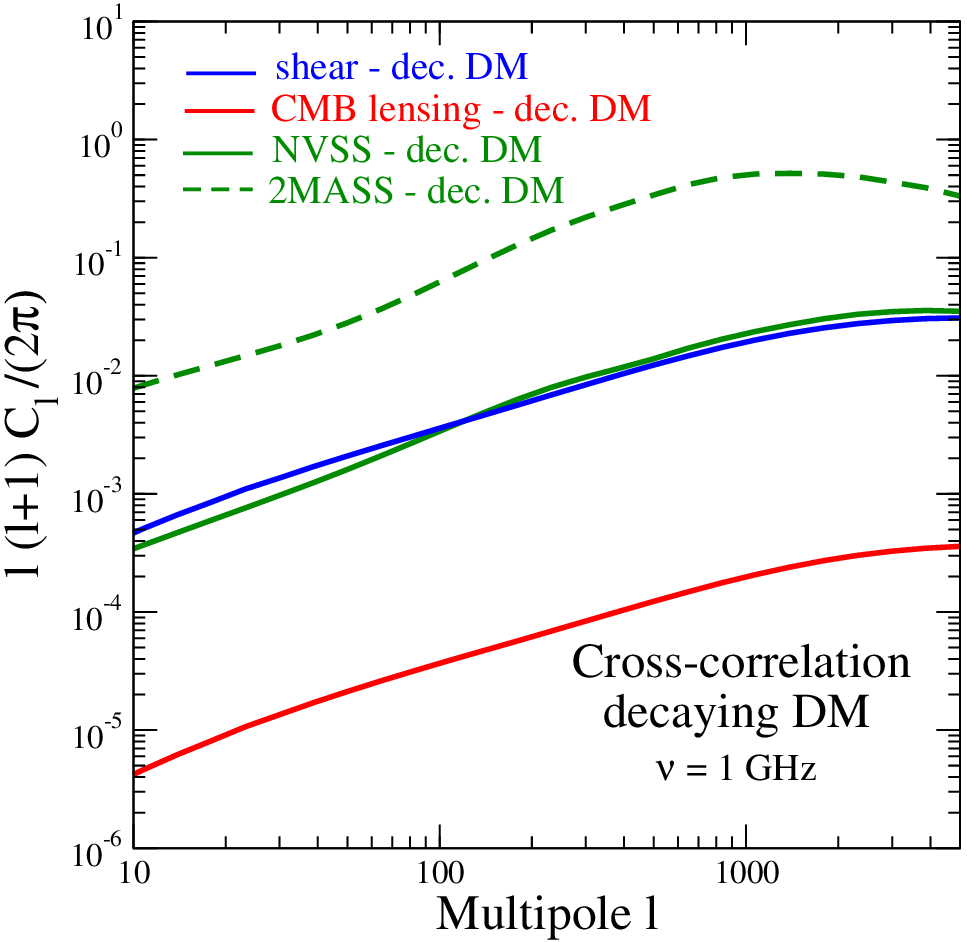}
\caption{Angular cross-correlation PS between the radio signals (at $\nu = 1$ GHz) and gravitational tracers (cosmic shear, CMB lensing and LSS tracers), for annihilating DM (left panel) and decaying DM (right panel). The computation is performed using the 3D PS models reported in Fig.~\ref{fig:ps2} and the window functions shown in Fig.~\ref{fig:winfun}. 
In the left panel, we also show (with thinner lines) the case of a cored DM distribution (as a possible results of particle diffusion).
}
\label{fig:corr3}
\end{figure}

The angular power spectra of annihilating DM are presented in the left panels of
Fig. \ref{fig:corr1},\ref{fig:corr2},\ref{fig:corr3}, while the decaying cases are reported in the right panels of the same figures. Fig. \ref{fig:corr1} shows the auto- and cross-correlations of the electromagnetic signals. The most anisotropic cases involve the radio emission. This is because the radio window function is peaked at low-redshift: the signal therefore comes from a relatively low number of halos and thus is quite anisotropic. The flatter window function of X-rays implies that in this case 
the number of objects contributing to the intensity field is larger, thus making the emission smoother
and the ensuing PS smaller. The gamma-rays case is intermediate to radio and X-rays: as shown
by its window function in Fig. \ref{fig:winfun}, the gamma-rays emission is peaked at low-to-intermediate redshifts and this implies a relatively larger anisotropy in the intensity, as compared to X-rays. We also notice that the radio PS flattens more rapidly at $\ell>10^3$ with respect to gamma and X-ray PS. 
The 1-halo term takes over at large multipoles and leads to constant $C_\ell$ (Poisson-noise) when sources are point-like, while instead it leads to a decreasing $C_\ell$ when structures start to be resolved. The flattening is therefore again related to the redshift distribution. Indeed the inner structure of closer objects, which are more important in the radio case, is resolved at larger angular scale with respect to more distant objects which contribute to the bulk of the emission in the gamma and X-ray cases.
We finally notice that at small angular scales, annihilating DM provides much more anisotropy
as compared to decaying DM, due to its dependence on $\rho^2$ and to the subhalo scheme adopted.

As mentioned above, we are taking an approximate description for radiative emissions (both IC and synchrotron) where all the power is radiated at the same place of $e^+/e^-$ injection, and therefore the density field of the emission is proportional to the DM density (for decaying DM) or DM density squared (for annihilating DM). One of the neglected effects is the diffusion of $e^+/e^-$, which might lead to a smoothing of the cuspy behaviour of the density profile into a more cored profile (for a more extended discussion on the impact of this approximation on radio anisotropies, see Ref. \cite{Fornengo:2011xk}). If we still assume that all the power is radiated within the object, we can model the effect of diffusion 
on the final $e^+/e^-$ distribution by replacing the NFW profile $\rho_{\rm NFW}\propto x^{-1}\,(1+x)^{-2}$ (where $x=r/r_s$ and $r_s$ is the scale radius) with a cored density, like e.g., $\rho_{\rm cored}\propto (1+x)^{-1}\,(1+x^2)^{-1}$.
The thick and thin dotted lines in the left panel of Fig.~\ref{fig:corr1} show the case with an NFW and cored distribution, respectively.
The difference is not dramatic (notice that this can also be seen as an estimate for the case where the DM distribution itself is cored, although the power spectrum we have been using is based on results of N-body simulations, thus fully consistent only with a cuspy profile).

Summarizing the behavior of the auto- and cross-correlation power spectra of the electromagnetic
signals among themselves, we can state that the radio emission exhibits the strongest anisotropy, both among the auto-correlation signals and in combination with the other emissions. There
are almost two orders of magnitude between the radio/radio auto-correlation signal as compared
to the X-rays/X-rays auto-correlation (which exhibit also the smallest among all the correlations
of electromagnetic signals) and one order of magnitude larger than the gamma/gamma autocorrelation. Interesting prospects are present for the cross-correlation radio/gamma, which is a factor of 3-5 larger than the gamma/gamma case. These general features occur both for the annihilating and decaying DM signals, with more power at large multipoles for annihilating DM as compared to the decaying case,
and with slightly more separation among the different cases again for annihilating DM as compared
to decaying DM.

Since we are dealing here with angular power spectra normalized to the average intensity, the actual
feasibility of detection will depend also on the absolute normalization level accessible by the different
detectors, which is detector specific. Being concerned in this paper with the theoretical properties
of the auto- and cross-correlation signals and with the assessment of their mutual impact,
we are not adopting here any specific experimental figure of merit: results shown in Fig. \ref{fig:corr1},\ref{fig:corr2},\ref{fig:corr3} can then be folded with the individual detector capabilities. In fact,
the experimental ability to disentangle an anisotropy signal will also depend on the specific features of the detectors and on the astrophysical backgrounds, which also produce an anisotropic
electromagnetic emission. Photon detectors at different wavelengths have intrinsically different
angular resolution: e.g. radio telescopes can resolve very fine details, and have an angular resolution
which is much better than gamma-rays detectors. They will therefore be a suitable instrument for
studying large multipoles $\ell$. On the other hand, at lower multipoles gamma-rays detector may be
more suitable than interferometric radio telescopes. The combination of the information coming from the auto-correlation
signal at different wavelengths, as well as cross-correlations of different signals, may therefore be a relevant tool to identify and characterize a DM signal. Concerning astrophysical backgrounds, a large
number of electromagnetic emitters are present, like e.g. active galactic nuclei, star forming galaxies,
etc.: their emission is anisotropic as well. The physical location of these astrophysical sources is
to some extent correlated to the DM structures: nevertheless, they possess different properties both
in redshift distribution (relevant for the ensuing window functions) and in spectral features. These
differences can be potentially used to attempt a separation between these backgrounds and the
DM signals discussed in this paper. Examples of the impact of astrophysical background have
discussed in the case of the gamma-rays auto-correlation \cite{Ando:2006cr,Taoso:2008qz,Xia:2011ax,Hensley:2012xj,Ripken:2012db,Ando:2013ff}, of the radio auto-correlation \cite{Zhang:2008rs,Fornengo:2011xk}
and of the cross-correlations gamma/shear \cite{Camera:2012cj}.  In this paper we are concerned on assessing the size and the relative impact of DM signals. A detailed analysis of the background sources is beyond the scope of the present paper, and for the cases for which studies are present we refer to the quoted references.

The cross-correlation of the gamma and radio emissions with gravitational tracers is shown in Figs.~\ref{fig:corr2} and \ref{fig:corr3}, respectively. For definiteness, also here we have considered
the gamma-rays emission at $E=1$ GeV and the radio emission art $\nu=1$ GHz. We notice that
the strongest correlation occurs with the 2MASS population, while the lowest is with the CMB lensing. 
This can be again understood by looking at the redshift dependence of the window functions.
To have a good overlapping with the electromagnetic DM source, the gravitational tracer has to be  peaked at relatively low redshift. Moreover, as we already discussed, the closer is the emission the more anisotropic it appears (and the ensuing angular PS flattens sooner, as a function of the multiple $l$).
From the window function behavior of Fig. \ref{fig:winfun}, we see that the radio signal and the
2MASS tracer are both strongly peaked at very low redshifts: this fact enhances the cross-correlation
both because of the large overlap and because closer sources are fewer and therefore more anisotropic.
This is clearly seen in the cross-correlations results of Fig.  \ref{fig:corr3}, where the angular PS
for radio/2MASS is significantly large. In the left panel of Fig.  \ref{fig:corr3}
we show again the effect of a cored distribution, which is even milder than for the auto-correlation PS (this occurs because this effect does not affect the LSS tracers and so now modifies only one of the two fields entering in the PS computation). Also the gamma/2MASS cross-correlation is relatively large, although smaller than the for the radio case. This is again due to the good overlap of
the gamma-rays and 2MASS window functions, nevertheless to a lower extent than in the radio case.

The cross-correlations with cosmic shear (originally proposed for gamma/shear in Ref. \cite{Camera:2012cj}) and with NVSS have very similar angular PS, which could potentially offer a tool to disentangle
a DM signal from astrophysical backgrounds. 
In the case of cross-correlation with the cosmic-shear, a tomographic approach is feasible, and represents an unique opportunity to test the different redshift scaling we have been discussing \cite{wip}.

The cross-correlation with the CBM lensing observable
is the smallest, due to the fact that CMB-lensing sources are distributed in a much deeper interval of  redshift, as compared to the electromagnetic emitters. We finally notice that
the decaying DM cross-correlation angular PS (but this applies also for the auto-correlation case),
even though they have a behavior similar to the annihilating case, nevertheless exhibit a
slower growth of $\ell(\ell+1)\,C_\ell$, which means less power at small scales: this can be understood 
by comparing the annihilating and decaying 3D power spectra in Fig.~\ref{fig:ps2} at $k\gtrsim 1\, {\rm Mpc}^{-1}$.

Summarizing the behavior of the cross-correlation power spectra of the electromagnetic
signals with the gravitational tracers, we can state that again the radio emission exhibits the strongest anisotropy signal. 
The largest effect occurs (both for radio and gamma-rays) in the cross-correlation
to the low-redshift 2MASS population: the angular PS spectrum in this case is about one order of magnitude stronger than the cross-correlations with cosmic shear and NVSS, and about three orders
of magnitude larger than the cross-correlation with the CMB-lensing observable.
We caution again that however such differences do not straightforwardly translate into the actual experimental capabilities.

\section{Conclusions}

In this paper we have discussed extragalactic anisotropies in the electromagnetic emission
produced by DM annihilation or decay as a promising tool to search for a DM signal.
We have first reviewed the formalism needed to compute a generic 2-point angular power spectrum by following the halo-model description of clustering of structures in the Universe.
This formalism was then applied to realize estimates involving relevant emissions induced by particle DM annihilations or decays.
We have discussed the features and the relative size of the various auto- and cross-correlation angular power spectra that can be envisaged for anisotropy studies. 

From the side of DM signals we have considered the full multi-wavelength
spectrum, including the synchrotron emission at radio frequencies, the IC radiation in the X-ray and gamma-ray bands, as well as the prompt emission of gamma-rays. 
The angular power spectra of auto-correlation of each of these signals and of the cross-correlation
between any pair of them is presented. 

As a way to enhance the capability of detection of such non-gravitational signals of DM (and to improve their disentanglement from other astrophysical backgrounds) 
we introduce their cross-correlation with maps tracing the gravitational potential.
We have analyzed this possibility studying specific gravitational tracers of DM distribution in the Universe: weak-lensing cosmic shear, large scale structure matter distribution and CMB-lensing. 

We have shown that cross-correlating a multi-wavelength DM signal
(which is a direct manifestation of its particle physics nature) with a gravitational tracer (which is a manifestation of the presence of large amounts of unseen matter in the Universe) may offer a prime
tool to demonstrate that what we call DM is indeed formed by an elementary particle.

\section{Acknowledgments}
Work supported by the INFN research grant {\sl Astroparticle Physics Project -- FA51}
and by the  {\sl Strategic Research Grant: Origin and Detection of Galactic and Extragalactic Cosmic Rays} funded by Torino University and Compagnia di San Paolo. N.F. acknowledges support of the
spanish MICINN Consolider Ingenio 2010 Programme MULTIDARK
CSD2009--00064.

\end{document}